# The analogy between optical pulse compression and optical coherence transformation

M.V. Lebedev

Institute of Solid State Physics, RAS, 142432, Chernogolovka, Moscow distr.

**Abstract**
A new type of an optical interferometer is discussed the phase difference between the interfering beams in which is substantially wavelength dependent. It is shown that the function measured with this device is an integral transform of the field correlation function. Possible applications for data encoding and spectral linewidth measurements are considered.

## 1. Introduction

The development of quantum information processing and computation demands further detailed consideration of individual photon emission and absorption processes, examination of qubit ensembles and new methods of information encryption. This needs detailed understanding of optical coherence and its dependence on the properties of the device used for coherence measurements (the interferometer). This dependence is usually supposed to be small, because phase shifts introduced by the optical components of the interferometer can be neglected. An example of Gires – Tournois (or step-phase Michelson) interferometer [1] and the interferometer analyzed below show that optical interferometry can give more detailed knowledge of light.

## 2. Optical scheme

Let us consider the scheme of an optical interferometer first proposed in [2] (see figure 1a). A plane monochromatic light wave falls onto the diffraction grating **G** at the angle $\psi$. After diffraction at the grating (diffraction angle $\varphi$) and reflection from the plane mirror **M** oriented parallel to the grating at distance **h** from it light falls onto the grating **G** again and after the second diffraction propagates parallel to the beam specularly reflected from the grating. One can observe far field optical interference between beams 0 and 1 in the focal plane of an objective placed after the interferometer (not shown in the picture). This scheme resembles that of an optical pulse compressor introduced by Treacy [3] (see figure 1b) and widely used in laser optics. This similarity became evident if one considers the images of the grating $G_1$ and beam trajectories formed with the mirror **M** (see figure 1c). Detailed theory of this interferometer was given in [4].

## 3. General consideration

It is obvious that the phase difference between interfering beams will depend substantially on wavelength in this device, because the optical path of the diffracted beam will change with the change of the wavelength. One can conclude that the resultant interference pattern will depend not only on the field correlation function as in a usual two beam interferometer but on the device properties as well. The interference pattern $I$ given with a two-beam interferometer can be expressed through the intensities of the interfering beams $I_1$ and $I_2$ in the usual way as:

$$I = I_1 + I_2 + 2\sqrt{I_1 I_2}\, J \tag{1}$$

Theoretical considerations [4] show that the interference term is proportional to:

$$J = Re\left\{ exp(-i\Delta_0) \int g_1(\tau) \Psi(\tau) d\tau \right\} \tag{2}$$

here $g_1(\tau)$ is the Green's function of the interferometer, $\tau$ - the time delay introduced by the interferometer, $\Delta_0$ - the carrier frequency phase shift of the quasimonochromatic light wave with frequency $\omega_0$ and $\Psi(\tau)$ - the envelope of the field correlation function. In the limit $g_1(\tau) = \delta(\tau - \tau_0)$ expression (2) gives the field correlation function itself as expected for an ordinary two beam interferometer which delays all spectral components of the incident light beam by the same delay time $\tau_0$.

Looking at (2) one can see that the interference term depends now not only on the field correlation function but also on the device properties concentrated in $g_1(\tau)$. If we limit ourselves by quasimonochromatic light waves ($\Delta\omega \langle\langle \omega_0$) one can use the approximate expression for the Green's function $g_1(\tau)$ [4]. Considering terms up to the second order only Eq. (2) can be transformed to:

$$J = Re\left\{ \left(\frac{i}{2\pi\Delta''}\right)^{1/2} exp(-i\Delta_0) \int exp\left[-i(\tau-\Delta')^2 / 2\Delta''\right] \Psi(\tau) d\tau \right\} \tag{3}$$

were $\Delta'$ and $\Delta''$ are the first and the second derivatives of the phase shift against frequency. If we compare this result with expression (14) of [3] which gives the complete waveform of the output pulse of the Treacy's pulse compressor we find that in appropriate variables the two simply coincide. The correspondence between variables is the following:

| Optical pulse compression | Optical coherence transformation |
|---|---|
| Time at the output plane $t'$ | First derivative of the phase shift $\Delta'$ |
| Time $t$ | Time delay $\tau$ |
| Carrier wave frequency $\omega_0$ | Carrier wave frequency $\omega_0$ |
| Carrier wave phase shift $\omega_0(t'+\tau_0) - \varphi_0$ | Carrier wave phase shift $\Delta_0$ |
| Delay time frequency derivative $-\dfrac{1}{\mu}$ | Second derivative of the phase shift $\Delta''$ |
| Input pulse waveform $A(t)e^{i\psi(t)}e^{-i\omega_0 t}$ | First order field correlation function of the input beam $A(\tau)e^{i\chi(\tau)}e^{-i\omega_0\tau} = \Psi(\tau)e^{-i\omega_0\tau}$ |

## 4. The analogy with pulse compression

This analogy has a deep physical origin based on the equivalence between time domain and frequency domain description of light propagation through the interferometer. In the frequency language we consider a monochromatic wave propagating through the system and the phase shift of this wave $\Delta(\omega)$ caused by such propagation. On the other hand, one can equivalently consider a wave packet propagation and its time delay $\tau$ caused by the interferometer. The disappearing of interference fringes is explained in the first language as an overlap of constructive and destructive interference of a large number of Fourier components of the quasimonochromatic optical wave with spectral width $\delta\omega$. The lost of coherence in the second language is treated as the vanishing of the overlap between the two pulses with spectral width $\delta\omega$ due to the increase of the length difference of their paths trough the interferometer. The first explanation seems to be more adequate if one works with cw light sources, while the second one is usual for pulsed laser light. The difference between a quasimonochromatic wave with spectral width $\delta\omega$ and a pulse with spectral width $\delta\omega$ is the existence of phase matching between different Fourier components in a latter case and the absence of this matching in the first one. This means that phase matching does not affect the first order interference. Mathematically this fact can be expressed in the following way. Treacy's gratings pair introduces a frequency dependent phase shift of the Fourier components $F_{in}(\omega)$ of the incoming pulse, so that the spectrum of the output pulse becomes:

$$F_{out}(\omega) = F_{in}(\omega) exp\left[i\left(\varphi_0 - \omega_0\tau_0 - \frac{\omega^2}{2\mu}\right)\right] \tag{4}$$

This frequency dependent phase shift is responsible for the effect of "pulse compression"- the reduction (or increase) of the pulse envelope duration. But the pulse autocorrelation function which gives also the visibility of interference fringes in interference experiments is independent of phase shifts:

$$\langle E(t)E^*(t+\tau)\rangle = \frac{1}{2\pi}\int |F(\omega)|^2 \exp(i\omega\tau)d\omega \tag{5}$$

For observables depending on the field correlation function only both frequency and time domain languages give identical results, that's why we can apply some well known results of the pulse compression technique directly to optical coherence transformation. But first of all we have to consider some restrictions on coherence transformation resulting from the stationarity of the incoming light beam. If we suppose the initial light beam to be a complex stationary random process we can use the well known property of its correlation function $\Psi(\tau) = A(\tau)e^{i\chi(\tau)}$:

$$\Psi(-\tau) = \Psi^*(\tau) \tag{6}$$

This gives for its amplitude and phase functions which are both real:

$$A(-\tau) = A(\tau) \text{ and } \chi(-\tau) = -\chi(\tau) \tag{7}$$

Tracy's gratings pair can effectively compress optical pulses with a positive frequency chirp, that is with a phase function:

$$\psi(t) = -\frac{1}{2}\mu t^2 \text{ with } \mu\rangle 0 \tag{8}$$

But such phase function is forbidden for a complex stationary random process as one easily can see from (7). The allowed function is:

$$\psi(t) = -\frac{1}{2}\mu t^2 \, sgn(t) \quad \mu\rangle 0 \tag{9}$$

The last relation means that the autocorrelation function of a stationary light beam cannot be effectively "compressed". That is the function of the mutual coherence $J(\Delta', \Delta'')$ of the two light beams produced by the interferometer cannot be made considerably shorter than the original autocorrelation function, because the "compression" of the leading part of the pulse is always accompanied with the "stretching" of its trailing one.

The autocorrelation function without phase modulation $\chi(\tau) \equiv 0$ is in some sense analogous to a spectral limited optical pulse, because $J(\Delta', \Delta'')$ cannot be "compressed" at all. But to "stretch" this function is possible. In this case the "mutual coherence length" of the two light beams can be appreciably increased. It should be noted that the autocorrelation function of every output light beam in the interferometer remains the same as that of the original one, because their frequencies are unaffected by the interferometer. This means that information can be encoded in this mutual coherence function $J(\Delta', \Delta'')$ and cannot be extracted with usual interference experiments from the beams taken alone.

One can see from (3) that the interference term depends on two parameters: $\Delta'$ and $\Delta''$. The first one is the time delay between the beams and it is present in every two-beam interferometer, whereas the second is negligible in traditional schemes. This two parameters can be varied independently, because addition of a constant to $\Delta'$ does not affect $\Delta''$. A possible optical solution is schematically shown in figure 1d.

## 5. Lineshape analysis of spectral lines

An independent variation of $\Delta''$ can be used in lineshape analysis of spectral lines in a following way. Let us consider a well known case of a radiating gas with Doppler broadened optical transition with a Lorenz spectrum. The field correlation function takes in this case a well known form:

$$\Psi(\tau) = exp\left[-\Gamma|\tau| - \frac{(\Delta\omega_D \tau)^2}{4} - i\delta\tau\right] \quad (10)$$

with $A(\tau) = exp\left[-\Gamma|\tau| - \frac{(\Delta\omega_D \tau)^2}{4}\right]$ and $\chi(\tau) = -\delta\tau$. This function is analogous to a symmetrical pulse with linear phase modulation. $\Gamma$ is here the radiative damping of the optical transition, $\delta$ - radiative frequency shift and $\Delta\omega_D$ - the Doppler broadening. The interference term takes the form:

$$J = Re\left\{\left(\frac{i}{2\pi\Delta''}\right)^{1/2} e^{-i\Delta_0} \int_{-\infty}^{+\infty} exp\left[\frac{-i(\tau - \Delta')^2}{2\Delta''} - \Gamma|\tau| - \frac{(\Delta\omega_D \tau)^2}{4} - i\delta\tau\right]d\tau\right\} \quad (11)$$

The integral can be taken analytically (see [5]) and the result reads:

$$J = Re\left\{\left(\frac{i}{2\Delta''}\right)^{1/2} \sqrt{\beta}\, exp\left[-i\left(\Delta_0 + \frac{\Delta'^2}{2\Delta''}\right)\right]\left(e^{\beta\gamma_+^2}[1 - \Phi(\gamma_+\sqrt{\beta})] + e^{\beta\gamma_-^2}[1 - \Phi(\gamma_-\sqrt{\beta})]\right)\right\} \quad (12)$$

$\Phi(z)$ is here the probability integral, $\beta = \dfrac{\Delta''}{\Delta\omega_D^2 \Delta'' + 2i}$, $\gamma_- = \Gamma - \left(\dfrac{\Delta'}{\Delta''} - \delta\right)i$ and

$\gamma_+ = \Gamma + \left(\dfrac{\Delta'}{\Delta''} - \delta\right)i$.

Suppose for a moment that $\delta = 0$. In this case $J(\Delta')$ becomes symmetrical against the point $\Delta' = 0$. This can be seen from the fact that $\gamma_+ \Rightarrow \gamma_-$ and vice versa when $\Delta' \Rightarrow -\Delta'$. The latter condition means that $J(\Delta')$ reaches its maximum at $\Delta' = 0$ which is also quite natural from a physical meaning of coherence, because no additional phase shift, caused by different beam delays is introduced in this case. Introducing a finite $\delta \neq 0$ shifts this curve as a whole along the $\Delta'$ axis. Experimental determination of this shift makes possible direct determination of the Lamb's shift from interference measurements. This result should be compared with the field correlation function (10) measured with an ordinary two-beam interferometer. There is no any shift of the interference pattern in the latter case. For the visibility maximum $\frac{\Delta'}{\Delta''} - \delta = 0$ and we get from (12):

$$J_0 = Re\left\{ \left(\frac{2i\beta}{\Delta''}\right)^{1/2} e^{-i\Delta_0} \exp(\beta\Gamma^2)\left[1 - \Phi(\Gamma\sqrt{\beta})\right] \right\} \tag{13}$$

It is interesting to consider some simple cases before analyzing the general expression (12). If we take $\Gamma$ to be negligible small, Eq.(11) coincides with Fresnel diffraction integral for the Gaussian light beam (the mathematical analogy between pulse compression and Fresnel diffraction was already mentioned by Treacy [3]), the probability integrals in square brackets compensate each other (because $\gamma_- = -\gamma_+$) and we get:

$$J = Re\left\{ \left(\frac{2i}{\Delta\omega_D^2 \Delta'' + 2i}\right)^{1/2} \exp\left[-i\Delta_0 - \frac{\Delta\omega_D^2}{(\Delta\omega_D^2 \Delta'')^2 + 4}\left(1 + \frac{\Delta\omega_D^2 \Delta''}{2}i\right)(\Delta')^2\right] \right\} \tag{14}$$

We see that interference visibility vanishes as in a usual two beam interferometer following a Gaussian function but the interference pattern exhibits now additional phase modulation quadratic in $\Delta'$. It was observed experimentally in [2] while varying the central wavelength of the Gaussian beam entering the interferometer. In the limit $\Delta\omega_D^2 \Delta'' \langle\langle 1$ the phase of interference pattern (14) becomes:

$$-i\Delta_0 - \frac{i}{8}\Delta\omega_D^4 \Delta''(\Delta')^2 \tag{15}$$

The second term reduces the number of interference maxima passed when varying the central wavelength of the Gaussian beam, because $\Delta'' \langle 0$. This phase modulation became considerable if $\Delta\omega_D^2 \Delta'' \propto 1$. Its practical use may be to reduce phase fluctuations of the output beam originating from the frequency instability of the input one.

In the limit $\Delta\omega_D^2 \to 0$ we arrive at a very interesting case of single atom radiation coherence transformation. We get:

$$J = Re\left\{\frac{1}{2} e^{-i\left(\Delta_0 + \Delta'\delta + \frac{\Delta''}{2}(\Gamma^2 - \delta^2)\right)}\left[e^{\Gamma(\Delta' - \delta\Delta'')}\left(1 - \Phi\left(\gamma_+\sqrt{\frac{\Delta''}{2i}}\right)\right) + e^{-\Gamma(\Delta' - \delta\Delta'')}\left(1 - \Phi\left(\gamma_-\sqrt{\frac{\Delta''}{2i}}\right)\right)\right]\right\} \tag{16}$$

The visibility maximum at $\Delta' = \delta\Delta''$ is:

$$J_{max} = Re\left\{ e^{-i\left(\Delta_0 + \frac{\Delta''}{2}(\Gamma^2 + \delta^2)\right)}\left[1 - \Phi\left(\Gamma\sqrt{\frac{\Delta''}{2i}}\right)\right] \right\} \tag{17}$$

The result of a numerical calculation of (16) is shown in figure 2. The interference pattern is determined with the exact form of a single photon wave packet including terms with phase modulation. The initial form of the field correlation function used in these calculations is in fact a result of theoretical consideration of single atom spontaneous radiative decay made in a frame of a certain simplified model. The emitted quantum remains, strictly speaking, entangled with the atom during the radiation process, but disentanglement can take place due to noisy environment. The precise solution of this problem is rather complicated and does not exist yet [6]. Experimental measuring of single atom radiative decay interference pattern opens thus a way of experimental investigation of a single photon wave packet and the influence of the environmental noise.

From a mathematical point of view such interferometer makes an integral transform (3) on a single photon wave packet. This transform is obviously reversible, because the same integral represents such reversible operations as pulse compression and Fresnel diffraction. This means that we can find the initial form of a single photon wave packet by measuring the interference pattern and simply applying to this pattern the inverse integral transform.

We can now return to the analysis of a general case (12). It is more or less clear that at small $\Delta'$ the interference pattern will be more sensitive to radiative broadening $\Gamma$, while at large $\Delta'$ the central role will be played by $\Delta\omega_D$. Numerical calculations confirm this assumption. To illustrate the calculations let us consider the luminescence of a quantum dot ensemble. Suppose every quantum dot emits a Lorenzian spectral line with the width $\Gamma$, central frequency $\omega_i$ and these frequencies are randomly distributed around some central frequency $\omega_0$. The frequency distribution is supposed to be Gaussian with the width $\Delta\omega_D$. For the sake of simplicity we ignore the radiative frequency shift $\delta$ which is usually quite small. Under these assumptions we can use Eq.(12) with the following parameter values: $\Gamma = 10^{10} Hz$, $\Delta\omega_D = 10^{12} Hz$, $\delta = 0$ which are typical for quantum dots ensembles. The result of numerical calculations with these parameters is shown in figure 3. Figure 3a represents the amplitude envelope of the interference pattern for three different spacings between the grating G and mirror M. One can see that this envelope is rather sensitive to $h$. The amplitude falls considerably in the vicinity of the zero delay and an overall broadening of $J(\tau)$ is clearly visible. The consequence of $J(\tau)$ stretching is the appearance of interference in a region of $|\Delta'| \geq 4 ps$ where it is not usually observed at all. This region of time delays may be used for information encoding. It is interesting to mention that information is in this case encoded in phase shifts between interfering beams and could therefore not be extracted from the beams taken alone with any usual interference experiments. The higher order correlation functions of individual beams should on the other hand be affected by the interferometer. This means for example that the intensity correlation function of the beam diffracted by the grating taken alone will be changed. This opens a way for controlling intensity correlation functions of light beams. Figure 3b shows the amplitude of interference at the visibility maximum ($\Delta' = 0$) as a function of $\Delta''$ for two different values of $\Gamma$. We see that the order of $\Gamma$ can be evaluated from such measurements while $\Gamma$ remains much less than $\Delta\omega_D$.

**6. Some technical remarks**

Some technical remarks are appropriate at the end. If we look at Eq.(12) we can see that $J$ can be regarded not only as a function of $\Delta'$ and $\Delta''$, but as a function of $\frac{\Delta'}{\Delta''}$ and $\Delta''$ also. This gives certain advantages because $\frac{\Delta'}{\Delta''}$ is in the original scheme independent of $h$ [4]:

$$\frac{\Delta'}{\Delta''} = \frac{\cos\varphi + m\frac{\lambda}{d}tg\varphi}{-\frac{1}{2\pi\cos^3\varphi}\left(\frac{m\lambda}{d}\right)^2}\frac{c}{\lambda} \tag{18}$$

This gives the possibility of direct measuring the dependence $J(\Delta'')$ with $\frac{\Delta'}{\Delta''} = const$ by varying $h$. Moreover the main wavelength dependence of (18) is proportional to the change of the value:

$$\Phi = \frac{2h}{\lambda}\left(\cos\varphi + m\frac{\lambda}{d}tg\varphi\right) \tag{19}$$

which can be considered as a "group" phase shift along the trajectory of a propagating light beam. This is because the term:

$$-\frac{1}{2\pi\cos^3\varphi}\left(\frac{m\lambda}{d}\right)^2 = -\frac{1}{2\pi\cos^3\varphi}(\sin\psi + \sin\varphi)^2 \tag{20}$$

can be considered in the first approximation as wavelength independent, at least if $\sin\varphi \approx 0$. The "group" phase shift can be made wavelength independent for a special choice of $\varphi$ for given $\lambda$ and $d$ [2], which makes the interferometer in this case achromatic in some sense.

It is straightforward now, after explaining its physical origin, to get the "achromatic condition" for a more general case of figure 1d. The delay between the arms of the interferometer can be made independent of $h$ by making the additional delay of the zero order beam proportional to $\frac{2h}{c}$:

$$\frac{\tilde{\Delta}'}{\Delta''} = \frac{\Delta' + \alpha\frac{2h}{c}}{\Delta''} \tag{21}$$

Differentiating Eq.(21) against frequency gives:

$$\left(\frac{\Delta' + \alpha\frac{2h}{c}}{\Delta''}\right)' = 1 - \frac{\Delta' + \alpha\frac{2h}{c}}{(\Delta'')^2}\Delta''' \tag{22}$$

Setting this derivative to zero we get:

$$\frac{\Delta'}{\Delta''} + \alpha\frac{2h}{c}\frac{1}{\Delta''} = \frac{\Delta''}{\Delta'''} \tag{23}$$

Substituting into Eq.(23) $\Delta'$, $\Delta''$ and $\Delta'''$ from [4]:

$$\Delta' = \frac{2h}{c}\left(\cos\varphi + m\frac{\lambda}{d}tg\varphi\right) \tag{24}$$

$$\Delta'' = -\frac{2h\lambda}{c^2}\left(\frac{m\lambda}{d}\right)^2\frac{1}{2\pi\cos^3\varphi} \tag{25}$$

$$\Delta''' = \frac{2h\lambda^2}{c^3}\left(\frac{m\lambda}{d}\right)^2\frac{3}{4\pi^2\cos^5\varphi} \tag{26}$$

we finally get:

$$\alpha = \frac{1}{\cos\varphi}\left[\frac{1}{3}\left(\frac{m\lambda}{d}\right)^2 - \frac{m\lambda}{d}\sin\varphi - \cos^2\varphi\right] \tag{27}$$

For small $\varphi \approx 0$:

$$\alpha = \frac{1}{3}\left(\frac{m\lambda}{d}\right)^2 - 1 < 0 \text{ for } \frac{m\lambda}{d} \approx 1 \qquad (28)$$

The last condition means that for every given $m, \lambda, d, \varphi$ we can find the appropriate coefficient $\alpha$ which minimizes the "chromatism" of the interferometer.

One can see from the consideration above that high dispersion of the grating is very important. The dispersion rises considerably if working at large $\varphi \to 90^0$. At $\varphi = 70^0$ the dispersion is 25 times that of $\varphi \approx 0^0$ and at $\varphi = 84^0$ 1000 times. But working at large $\varphi$ needs taking into account correction to the Green's function proportional to $\Delta'''$ [4], because this correction grows rapidly with $\varphi$. Variation of $h$ gives in this case large displacement of the beams along the grating surface and large grating size is needed as a consequence. For example, the distance between the points of the beam intersections with the grating is for $\varphi = 70^0$ $l = 5,5h$ and for $\varphi = 84^0$ $l = 20h$. The use of two separate gratings is in this case more practical, because the central part of the grating is not used at all.

Theory of [4] and the above consideration as well is valid strictly speaking in the case only when the second zero order reflection of the light beam from the grating passes the mirror M. If this condition is violated multiple overlapping output beams occur (see figure 4). This gives for the validity of the consideration above the following condition:

$$2htg\varphi \geq D \qquad (29)$$

where $D$ is the size of the mirror M. The same condition holds for beams 0 and 1 to be well separated in space after the interferometer (we assume that the diffracted light fills the whole aperture of the mirror) so that the variant of figure 1d can be realized. The multiple beams case of fig4 is in some sense analogous to a Fabri–Perott interferometer. The appropriate consideration will be given in a separate paper.

**Acknowledgements**

The author would like to thank prof. O.V.Misochko for many helpful discussions and critical reading of the manuscript.

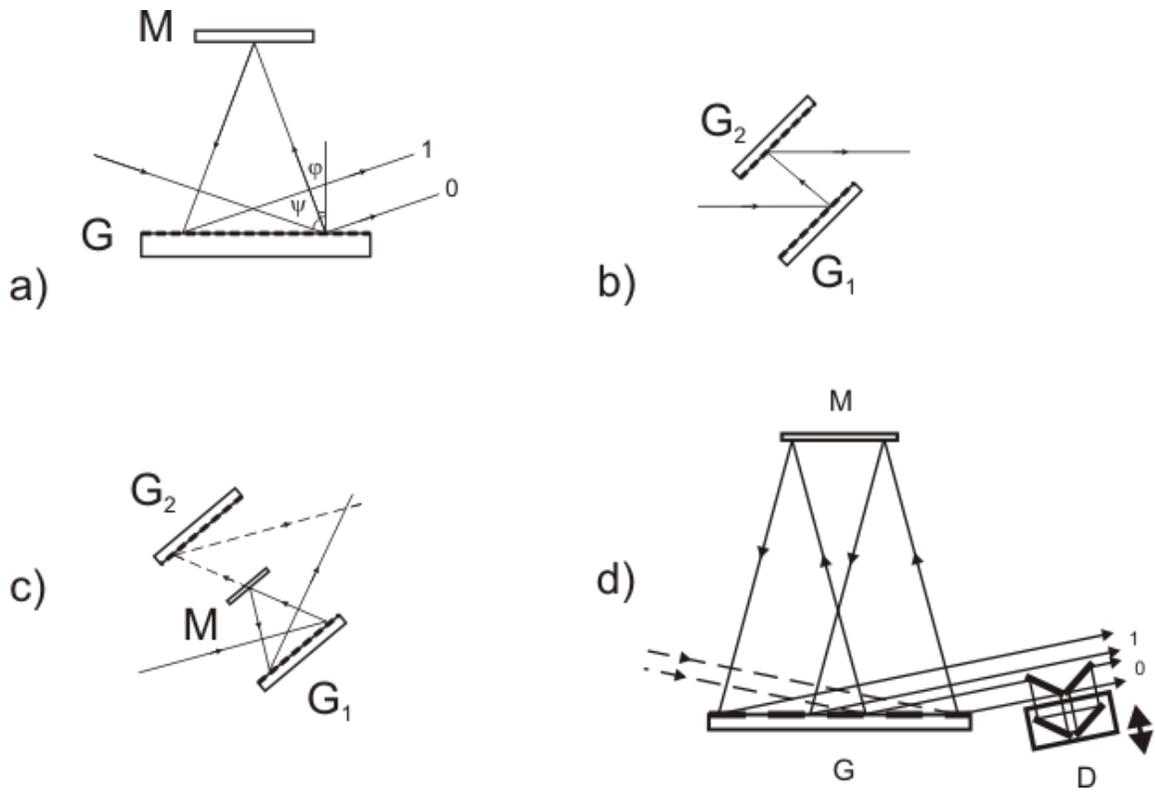

**Figure 1.** a) Optical scheme of the interferometer, b) Treacy's pulse compressor, c) The equivalence of diffracted beam path in Treacy's pulse compressor and the interferometer, d) A possible variation of the delay between interfering beams.

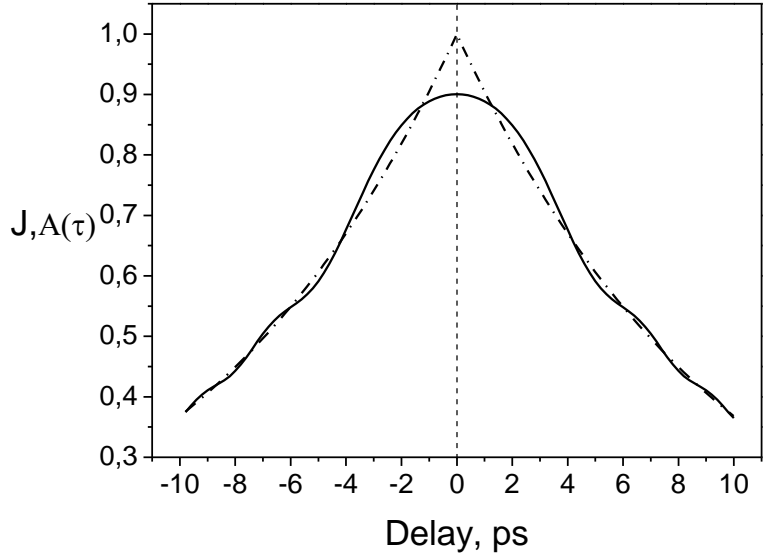

**Figure 2.** The comparison of the field correlation function amplitude envelope $A(\tau)$ (dot dashed) with the envelope of the interference pattern $J(\tau)$ for a single quantum dot luminescence. Grating 1200 mm$^{-1}$, m=1, $\varphi = 6^0$, $\psi = 37,38^0$, $d = 0,833 \mu m$, $\lambda = 0,589 \mu m$, $h = 300 cm$, $\Gamma = 10^{11} Hz$. The values of $\omega_D = 10^6 Hz$ and $\delta = 10^7 Hz$ were taken small but finite to avoid possible problems in numerical calculation procedure.

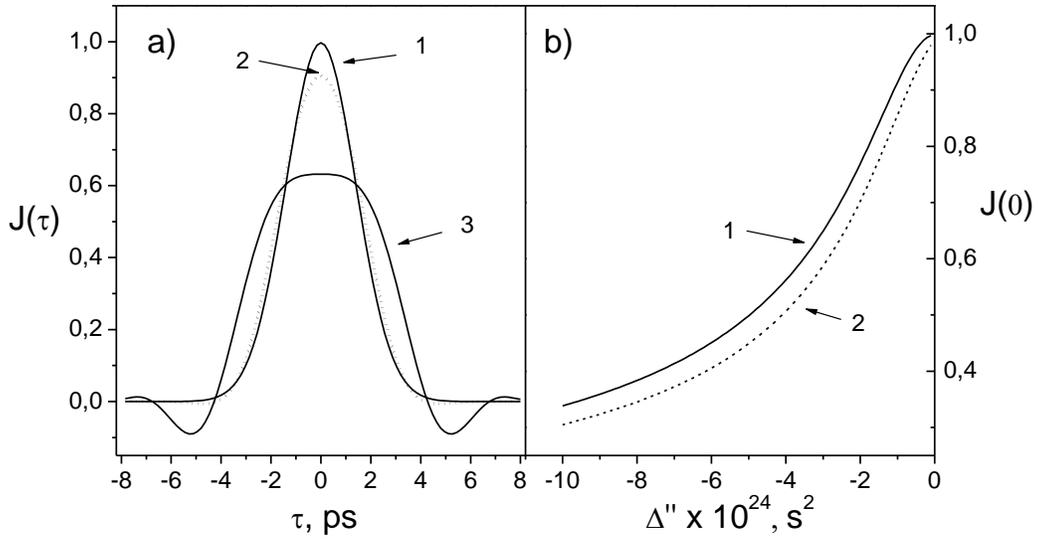

**Figure 3.** a) The envelope of the interference pattern $J(\tau)$ of the quantum dot ensemble luminescence for different spacings between the mirror and the grating: 1 - $h=10 cm$, 2- $h=100 cm$, 3 - $h=300 cm$, $\Gamma=10^{10} Hz$, $\omega_D=10^{12} Hz$. The reduction of the zero delay amplitude and considerable broadening of the interference pattern envelope is clearly seen. b) Zero delay amplitude of the interference pattern $J(0)$ as a function of $\Delta''$ for two different values of $\Gamma$: 1 - $\Gamma=10^{10} Hz$, 2 - $\Gamma=10^{11} Hz$. $\Delta''$ varies from $-10^{-25}$ s$^{-2}$ to $-10^{-23}$s$^{-2}$. This corresponds to variation of $h$ approximately from 10cm to 1000cm. All other parameters are the same as for figure 2. The dependence on $\Gamma$ is measurable, this opens a possibility of evaluating $\Gamma$ in the presence of a large inhomogeneous broadening of the spectral line.

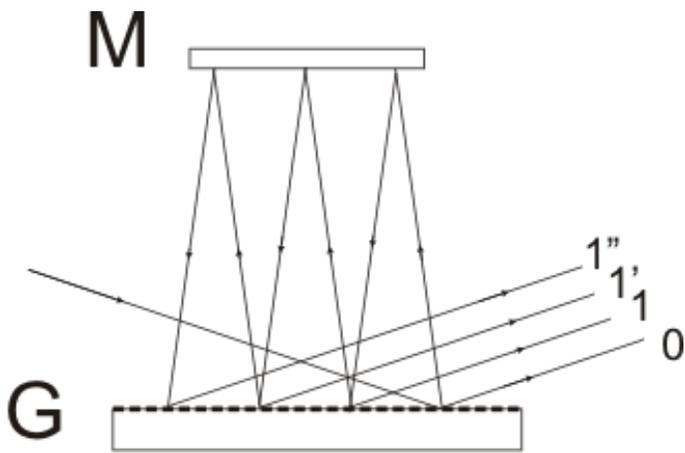

**Figure 4.** Possible multiple beam interference at small diffraction angles.